\documentclass[12pt]{iopart}

\usepackage{iopams}  
\usepackage{graphicx}
\usepackage{epstopdf}

\newcommand{\slfrac}[2]{\left.#1\middle/#2\right.}

\begin{document}

\title{Aberration-like cusped focusing in the post-paraxial Talbot effect}

\author{James D Ring$^{1}$, Jari Lindberg$^1$, Christopher J Howls$^2$ and Mark R Dennis$^1$}

\address{$^1$H. H. Wills Physics Laboratory, University of Bristol, Tyndall Avenue, Bristol, BS8 1TL, UK.}
\address{$^2$School of Mathematics, University of Southampton, Highfield, Southampton, SO17 1BJ, UK.}
\ead{james.ring@bristol.ac.uk}
\begin{abstract}
We present an analysis of self-imaging in a regime beyond the paraxial, where deviation from simple paraxial propagation causes apparent self-imaging aberrations.
The resulting structures are examples of aberration without rays and are described analytically using post-paraxial theory. 
They are shown to relate to, but surprisingly do not precisely replicate, a standard integral representation of a diffraction cusp.
Beyond the Talbot effect, this result is significant as it illustrates that the effect of aberration -- as manifested in the replacement of a perfect focus with a cusp-like pattern -- can occur as a consequence of improving the paraxial approximation, rather than due to imperfections in the optical system.
\end{abstract}

\pacs{42.25.Fx, 42.15.Fr, 42.25.Hz}

The Talbot effect \cite{Talbot:1836p771,Patorski:1989p1054} is a well-known self-imaging phenomenon in classical optics, where a transversely periodic field reproduces itself at periodic distances in the propagation direction.
As a natural consequence of paraxial wave propagation, it has also been observed in the diffraction of matter waves \cite{Chapman:1995p1897} and has an analogue in the revivals of quantum wavepackets \cite{Averbukh:1989p449}.
Recently, the Talbot effect for surface plasmons \cite{Dennis:1996p20,MartinezNiconoff:2008p382, Maradudin:2009,Zhang:2009p1,Cherukulappurath:2009p6} has attracted significant attention because of applications in areas such as nanolithography \cite{Wan:2010p380,Liu:2005p2127}.

The perfect image revival of the classical Talbot effect can be understood in terms of Fresnel diffraction \cite{Patorski:1989p1054} and interpreted as an artefact of the corresponding paraxial approximation.
Refining the propagation beyond the paraxial approach leads to aberration-like distortion of perfect self-imaging \cite{CohenSabban:1983p1601, Chang:2005p44,Leger:1990p2126}.
Nonparaxial self-imaging has been studied previously for binary gratings using both scalar theory \cite{Berry:1996p24,Mejias:1991} and full electromagnetic treatment \cite{Noponen:1993p495}, and has particular importance in plasmonics where the transverse periodicity is comparable to the plasmon wavelength \cite{Dennis:1996p20}. Such nonparaxial self-images are not perfect revivals as they are in the paraxial case, and are only approximately periodic in the direction of propagation.

In this paper, we investigate a nonparaxial Talbot effect and explain that the resulting apparent aberration -- the maximum intensity shifts and distorts to a cusp-like pattern -- is not caused by imperfection in a lens or optical system (since the Talbot effect is lensless imaging), but rather as an actual improvement of the approximation, appropriate to our diffraction regime.
Specifically, we invoke the fourth-order post-paraxial approximation \cite{CohenSabban:1983p1601, Leger:1990p2126, Berry:1996p24} to investigate the Talbot-based focusing of scalar waves in a regime which is nonparaxial, but not extremely so: the transverse periodicity is of the order of some tens of wavelengths.
This explains the appearance of diffraction cusps in the intensity patterns of surface plasmons calculated in Ref.~\cite{Dennis:1996p20}. 
We show that these aberration-like effects are not unique to plasmons and are connected to a standard integral representation of the diffraction pattern about a cusped caustic \cite{FNye:1999p2050} despite the absence here of any actual ray caustic.
Furthermore, we obtain a simple expression for the approximate location of the post-paraxial foci, which are shifted to be short of the paraxial Talbot distance.

To demonstrate the physics of the post-paraxial regime, we employ a general two-dimensional (2D) formalism (transverse in $x$ and propagating in $z$) which displays properties of both paraxial and nonparaxial regimes in different limits. 
Specifically, it consists of a transverse array (with period $a$ in the $x$-direction) of monochromatic 2D point sources of wavelength $\lambda$.
The interference of the propagating waves in the $x$-$z$ plane weaves a diffraction carpet, which is similar to that studied in the plasmon Talbot effect of \cite{Dennis:1996p20}. 
The paraxial approximation applies when $\gamma\equiv a/\lambda$ is asymptotically large. 
In this case the Talbot distance $2a^2/\lambda$ is the spacing between planes of image revival. 
%
%
\begin{figure}[h]
\centerline{
\includegraphics[width=15cm]{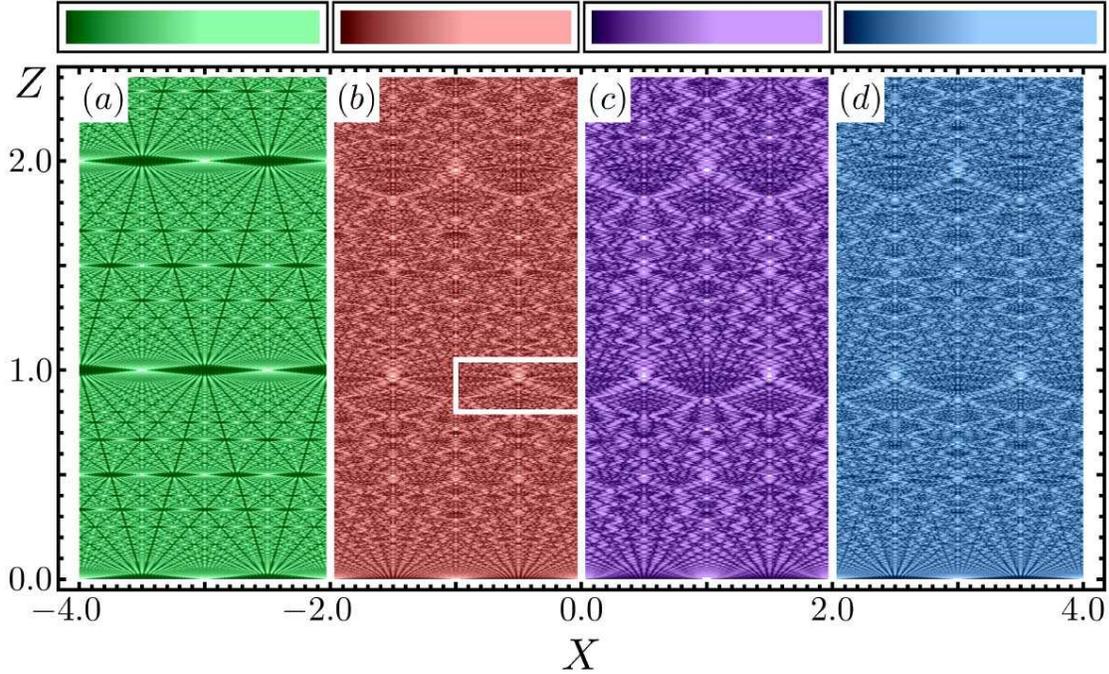}}
\caption{Intensity distributions ($\gamma=18$) for $(a)$ the paraxial field of Eq. \ref{eq:parax}; $(b)$ exact field of Eq.~\ref{eq:helm}; $(c)$ numerically simulated 1D binary gold grating (of refractive index $0.2+3.44\mathrm{i}$, thickness $100$ nm and slit width to grating period ratio of $0.05$) illuminated by a normally incident, TE-polarized, monochromatic ($\lambda = 632.8$ nm) plane wave; and $(d)$ the post-paraxial field as given by Eq. \ref{eq:ppp}. 
A value of $\gamma=18$ is outside the region of validity for the paraxial approximation, since $(a)$ is fundamentally different to $(b)$. The fields $(b)$, $(c)$ and $(d)$ exhibit foci (indicated by the white box) which are aberrated compared to $(a)$.}
\label{fig:regcomp}
\end{figure}
The initial pattern also revives at odd multiples of $a^2/\lambda$, albeit transversely shifted by $a/2$, leading to an alternate definition of the Talbot distance \cite{Berry:1996p24} of $z_\mathrm{T}\equiv a^2/\lambda$ (used in this paper). 
In what follows, we use the dimensionless variables $X=x/a$ and $Z=z/z_{\mathrm{T}}$.
In general, for any $\gamma$, the field at a distance $Z$, propagating according to the 2D Helmholtz equation, is given by (cf. Eq.~(1) of \cite{Dennis:1996p20})
%
%
\begin{eqnarray}
\psi(X,Z)= \sum_{n=-N}^{N} \mathrm{exp}\left[2\pi\mathrm{i}\left(nX+\gamma^2 Z\sqrt{1-n^2/\gamma^2} \right)\right]\label{eq:helm},
\end{eqnarray}
where we assume negligible attenuation and consider only propagating waves, i.e.,  $N= [\gamma]$ with square brackets denoting the integer part. Eq.~\ref{eq:helm} can be approximated by expanding
%
%
\begin{eqnarray}
\sqrt{1-n^2/\gamma^2}\approx 1-n^2/2\gamma^2-n^4/8\gamma^4.\label{eq:expansion}
\end{eqnarray}
In the paraxial approximation to Eq.~\ref{eq:helm}, the quartic term in Eq.~\ref{eq:expansion} is negligibly small, so 
%
%
\begin{eqnarray}
\psi_\mathrm{p}(X,Z) = \sum_{n=-N}^{N} \mathrm{exp}\left[\mathrm{i}\left(2\pi nX-\pi n^2 Z\right)\right]\label{eq:parax},
\end{eqnarray}
where we have neglected the overall phase factor of $\mathrm{exp}(2\pi\mathrm{i} z/\lambda)$.
%
%
%
%
\begin{figure}[h]
\centerline{
\includegraphics[width=15cm]{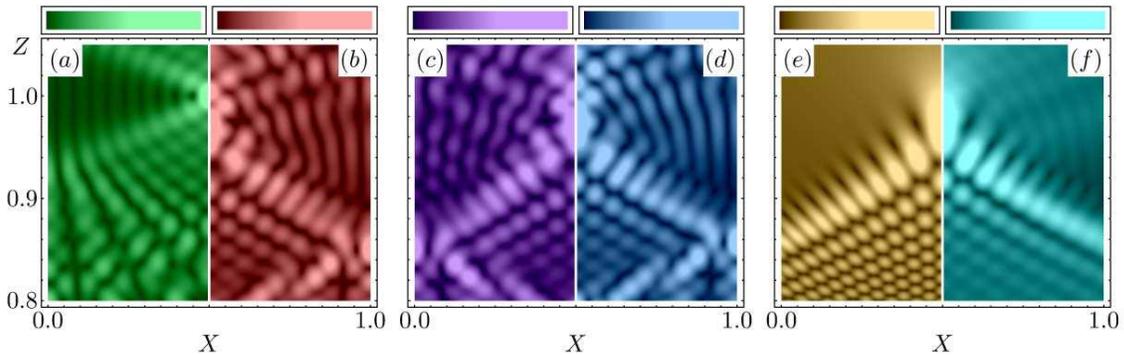}}
\caption{A comparison of foci symmetric about the centre of each image ($\gamma=18$, $M=9$) with colours corresponding to Fig.~\ref{fig:regcomp}; $(a)$ paraxial focus of Eq.~\ref{eq:parax}; $(b)$ the exact field of Eq.~\ref{eq:helm}; $(c)$ numerically simulated focus according to the parameters of Fig.~\ref{fig:regcomp}$(c)$; $(d)$ post-paraxial focus of Eq.~\ref{eq:PSFresult} ($|\mu|\le1$); $(e)$ the Pearcey function Eq.~\ref{eq:pe}; and $(f)$ the incomplete Pearcey function ($\mu=0$ term of Eq.~\ref{eq:PSFresult}).}
\label{fig:cusps}
\end{figure}
As shown in Fig.~\ref{fig:regcomp}$(a)$, the initial paraxial field revives for integer $Z$, since the second term in Eq.~\ref{eq:parax} will introduce a factor of $\pm1$, and we recover the same field as for $Z=0$. The factor of $\pm1$ determines whether or not the self-image is shifted by half a period. 
A post-paraxial refinement is made by retaining the quartic term in Eq.~\ref{eq:expansion} \cite{CohenSabban:1983p1601,Leger:1990p2126,Berry:1996p24}, giving
%
%
\begin{eqnarray}\label{eq:ppp}
\psi_{\mathrm{pp}}(X,Z)=\sum_{n=-N}^{N} \mathrm{exp}\left[\mathrm{i}\left(2\pi nX-\pi n^2 Z - \pi n^4 Z / 4\gamma^2\right)\right].
\end{eqnarray}

Comparison between Eq.~\ref{eq:ppp} and Eq.~\ref{eq:parax} shows that the perfect $Z$-periodicity of the Talbot effect is an artefact of the paraxial approximation, and is destroyed in the post-paraxial regime by the inclusion of the quartic (and possibly higher terms) \cite{Berry:1996p24}. 
Figs.~\ref{fig:regcomp}$(b)$ and \ref{fig:regcomp}$(d)$ show the exact and post-paraxial intensity distributions respectively, and it is immediately apparent that while the foci of these fields are similar to one another, they are very different, and less defined, from those of the paraxial case, ($a$).
Comparing the foci of each regime in Fig.~\ref{fig:regcomp} it is clear that the perfect foci of the paraxial approximation of $(a)$ have been perturbed in the post-paraxial regime to cusp-like distributions, evident in $(b)$ and $(d)$.
Fig.~\ref{fig:regcomp}$(c)$ shows a separate simulation for a binary grating, which has been calculated numerically using rigorous electromagnetic theory \cite{Noponen:1993p495, Moharam} (parameters given in the caption), which very closely resembles the exact and post-paraxial fields in $(b)$ and $(d)$, including the cusp-like foci.
While such a pattern is typical for aberration, its appearance here is because of a refinement of the usual Fresnel paraxial approximation.
In fact, the inclusion of a quartic term destroying a perfect focus is exactly analogous to the spherical aberration of a parabolic mirror, though remarkably, in Fig.~\ref{fig:regcomp} it is \textit{improvement} of the approximation and not any aberration in the system that causes the effect!

The fields in Fig.~\ref{fig:regcomp} are those for an infinite array, and applying the paraxial approximation to such a system is not fully consistent \cite{Mejias:1991}, since at all $Z$ there would be oblique contributions to the field from arbitrarily transversely distant sources -- a fact that is at odds with the concept of paraxiality.
Instead we model a finite array of $N$ sources by truncating the sums of Eqs.~\ref{eq:helm}, \ref{eq:parax} and \ref{eq:ppp} at $\pm M$,  where $M\equiv[N/2Z]$ \cite{Berry:1996p24,Berry:1988p1604}, embodying the walk-off effects of finite arrays \cite{Patorski:1989p1054}.
Fig.~\ref{fig:cusps} shows the various foci according to this truncation.
Comparing the paraxial focus Fig.~\ref{fig:cusps}$(a)$ to the exact Fig.~\ref{fig:cusps}$(b)$, the aberration effect is clear.
Fig.~\ref{fig:cusps}$(c)$ shows the cusp pattern from the diffraction grating of Fig.~\ref{fig:regcomp}$(c)$, but includes only diffraction orders up to $\pm M$.
There is a slight discrepancy between $(b)$ and $(c)$ because of the equal amplitudes of the planes waves in the superposition of Eq.~\ref{eq:helm} compared to varying amplitudes of the diffraction orders of the simulation of $(c)$.
The fields are nevertheless very similar and the cusp reveals the breakdown of the paraxial approximation.
Below, we develop an analytic post-paraxial approximation to these fields, shown in Figs.~\ref{fig:regcomp}$(d),$ \ref{fig:cusps}$(d),$ from inclusion of the quartic term (and no higher), which is enough to capture the structure of the exact and simulated foci.
Including higher terms improves the quantitative agreement, but without further qualitative changes.

To derive a simple representation of the complicated interference pattern, we follow a similar technique to Ref.~\cite{Berry:1996p24} but employ a \emph{finite} Poisson sum formula \cite{Civi:1999p688} to account for the truncation at $\pm M$ of the spectrum. 
Applying this to $\psi_{\mathrm{pp}}(X,Z)$ yields the main result of this paper,
\begin{eqnarray}
\psi_{\mathrm{pp}}(\xi, \zeta)&=\frac{f(-M; \xi, \zeta)}{2}+\frac{f(M; \xi, \zeta)}{2} + \sum_{\mu=-\infty}^\infty\int_{-\tau( \zeta)}^{\tau(\zeta)}\mathrm{d}s\ g(s; \xi, \zeta, \mu),\label{eq:PSFresult}
\end{eqnarray}
where $g(s; \xi, \zeta, \mu)\equiv h(\zeta)\mathrm{exp}\left\{-\mathrm{i}\left[s^4+\sigma_Z(\zeta)s^2+\sigma_X (\xi,\mu,\zeta)s\right]\right\}$, and
\begin{eqnarray}
f(n; \xi, \zeta)\equiv\mathrm{exp}\left\{\mathrm{i}\pi \left[2\xi n -\zeta n^2-(T+\zeta)n^4/4\gamma^2\right]\right\};\nonumber\\
h(\zeta)\equiv\sqrt{2\gamma}/[\pi(T+\zeta)]^{\frac{1}{4}};\nonumber\\
\tau(\zeta) \equiv M/h(\zeta);\nonumber\\ 
\sigma_X(\xi,\mu,\zeta)\equiv 2\pi(\xi-\mu) h(\zeta);\nonumber\\
\sigma_Z(\zeta)\equiv\pi h^2(\zeta)\zeta.\label{eq:scalings}
\end{eqnarray}
In the above, we have set $Z=T+\zeta$, where $\zeta$ is a small shift from the $T^\mathrm{th}$ Talbot distance, and $\xi=X-T/2$.
The terms $f(\pm M; \xi, \zeta)$ are end-point contributions from the truncation of the spectrum.
We recognise that the integral in Eq.~\ref{eq:PSFresult} is an incomplete version \cite{Nye:2003p882} of the Pearcey function \cite{Pearcey:1946p499,DLMF} (i.e. with a finite integration domain).
The Pearcey function is given by
%
%
\begin{eqnarray}\label{eq:pe}
\mathrm{Pe}(x,z)\equiv \int_{-\infty}^\infty \mathrm{d}s\ \mathrm{exp}\left[\mathrm{i}\left(s^4+zs^2+xs\right)\right],
\end{eqnarray}
which describes the generic diffraction pattern decorating a cusp caustic \cite{FNye:1999p2050,DLMF}.

The sum in Eq.~\ref{eq:PSFresult} represents an array of cusp-like foci in the $x$-direction, with transverse and longitudinal scalings given by $\sigma_X$ and $\sigma_Z$.
Although in optics it is usual to see the Pearcey function as decoration around a cusp caustic of rays that arise from a geometric optics approach \cite{FNye:1999p2050}, this is not necessary \cite{Berry:1999p335}.
Since here we are outside the applicability of geometrical optics, the propagating function Eq.~\ref{eq:helm} cannot be approximated in terms of rays, and instead the Pearcey function has come from the quartic post-paraxial term in Eq.~\ref{eq:ppp}.
Importantly, since the maximum of the Pearcey function is below the geometric focus, the maxima of the post-paraxial self-images are below the paraxial Talbot distance. 
This differs from the nonparaxial Talbot distance \cite{Zhang:2009p1,Noponen:1993p495} which is accurate when only a small number of diffraction orders contribute.

Fig.~\ref{fig:cusps}$(d)$ shows the diffraction pattern near the Talbot distance ($Z=1$) according to Eq.~\ref{eq:PSFresult} for $|\mu|\le1$. 
The intensity of the Pearcey function, scaled to the size set by $\sigma_X$ and $\sigma_Z$, is shown in Fig.~\ref{fig:cusps}$(e)$.
Fig.~\ref{fig:cusps}$(f)$ is the incomplete Pearcey function (the $\mu=0$ term of Eq.~\ref{eq:PSFresult}).
Comparing $(b)$ and $(d)$, we see that although the sum in Eq.~\ref{eq:PSFresult} has infinitely many terms, only the first few are needed to accurately describe the fine details of the field about the focus.
The transverse fringes come from the truncation of the integral in Eq.~\ref{eq:PSFresult}, and subsequently are absent from the Pearcey function, $(e)$, and are just evident in $(f)$.
In all these images, it is clear that the maximum intensity is short of the paraxial Talbot distance.
We now obtain an expression for the approximate location of the maximum intensity.

\begin{figure}
\centerline{
\includegraphics[width=11cm]{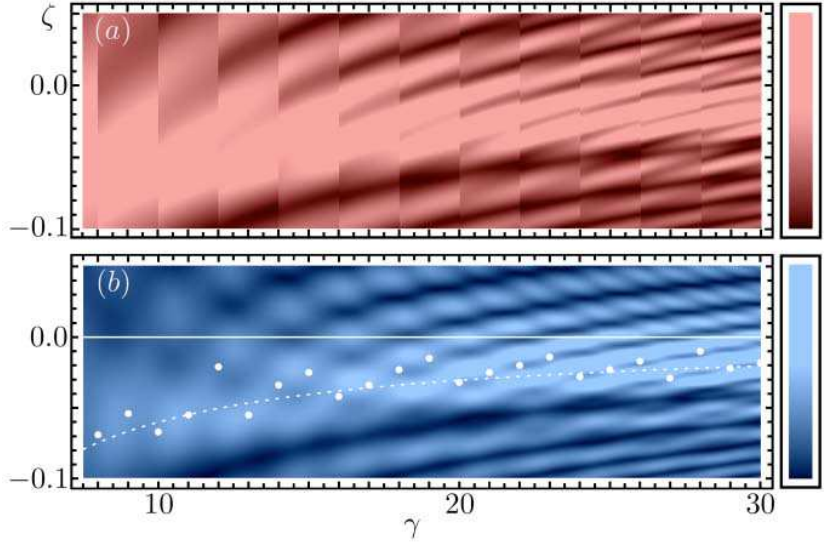}}
\caption{The intensity of $(a)$ the exact field, Eq.~\ref{eq:helm} (normalised by $(2[\gamma/2]+1)$) and $(b)$ the post-paraxial field (Eq.~\ref{eq:PSFresult}, summing $|\mu|\le1$) as a function of $\gamma$. The dashed line is the approximate location of maximum intensity according to Eq.~\ref{eq:peak}. The dots are the actual maxima of the exact field for integer $\gamma$. The solid line is the Talbot distance.}
\label{fig:feather}
\end{figure}

As a first approximation we take only the $\mu=0$ term of Eq.~\ref{eq:PSFresult} and let $\tau(\zeta)$ go to infinity.
This is equivalent to approximating the focal region near the first Talbot distance with the Pearcey function, Eq.~\ref{eq:pe}.
By using the numerically calculated, constant position of maximum intensity of the unscaled Pearcey function, $Z_{\mathrm{Pe}}\approx -2.199$, we can employ the scalings of Eq.~\ref{eq:scalings} to calculate the approximate location of the maximum intensity of the post-paraxial focus as a shift, $\zeta_{\mathrm{peak}}$, from the $T^{\mathrm{th}}$ Talbot distance.
By setting $\sigma_Z (\zeta_{\mathrm{peak}})=Z_{\mathrm{Pe}}$, we obtain
%
%
\begin{eqnarray}
\zeta_{\mathrm{peak}}=\slfrac{\left(Z_{\mathrm{Pe}}^2+Z_{\mathrm{Pe}}\sqrt{Z_{\mathrm{Pe}}^2+16\pi\gamma^2T}\right)} {8\pi\gamma^2}.\label{eq:peak}
\end{eqnarray}
Fig.~\ref{fig:feather}($a$) shows the intensity of the exact field along the symmetry axis of the cusp $\xi=0$ for $-0.1\le\zeta\le0.05$ as a function of increasing $\gamma$.
Asymptotically large $\gamma$ gives rise to maximum intensity at the Talbot distance ($\zeta=0$ in the paraxial limit), but for smaller $\gamma$ it is located short of this.
Fig.~\ref{fig:feather}$(b)$ shows the post-paraxial approximation of Eq.~\ref{eq:PSFresult} including only the terms $|\mu|\le1$.
The dashed curve is the location of the maximum intensity according to the approximation of Eq.~\ref{eq:peak}. 
Note that this depends on $T$, such that consecutive foci will be shifted from their paraxial counterparts by an increasing amount.
The dots indicate the maxima of the exact field (Eq.~\ref{eq:helm}) for integer $\gamma$, though they do not always correspond to well localised single peaks, as in the case of $\gamma=12$ and $\gamma=28$.
The intensity fringes that we observe in Fig.~\ref{fig:feather} are a result of interference between the $\mu=0$ and the neighbouring $\mu=\pm1$ terms.
Inclusion of more $\mu$-terms reconstructs the discontinuities of Fig.~\ref{fig:feather}$(a)$ (which arise from the limits of the sum in Eq.~\ref{eq:helm} changing discretely with $\gamma$).
%
%

To conclude, we have applied scalar post-paraxial theory to explain the observed cusp-like foci in the Talbot effect for the post-paraxial regime, which have replaced the point foci of paraxial theory.
While the analytic form of these post-paraxial foci is related to the standard description of diffraction about a cusp, there is a subtle distinction since it is the \emph{incomplete} Pearcey function that occurs and not the standard Pearcey function. 
The finite integration domain of the incomplete Pearcey is of significant asymptotic importance when compared with the infinite domain of the complete Pearcey function of Eq.~\ref{eq:pe}.
This is clear when comparing intensity patterns - while both are cusp-like distributions, there is additional complex detail accounted for in the post-paraxial Talbot effect that is absent from the normal Pearcey function.
Another important difference is that there is no ray caustic underpinning the structure of the post-paraxial cusp foci as there is with the complete Pearcey function.
Despite their differences, the location of the maximum of intensity of the post-paraxial foci can be fairly well approximated using a scaled Pearcey function.

It is curious to see that the effect of post-paraxiality on a focus appears to be that of aberration, when compared to the paraxial case.
However, while the analytic form mimics that for aberration (hence the inclusion of the Pearcey function, albeit incomplete), it actually embodies an improvement of the paraxial approximation for the post-paraxial regime. The phenomenon is a universal effect that occurs as the parameters of a system (regardless of its exact realisation) move away from paraxiality.

The results in this paper may plausibly find application in plasmonics, since it is a field in which the pertinent wavelengths and periodicities are comparable, and so nonparaxial effects are important. While the lengthscales considered here are slightly greater than those experimentally investigated so far, it would be both interesting and challenging to confirm the aberration-like effects of post-paraxiality on the Talbot effect.

\ack
We gratefully acknowledge the funding of the EPSRC. MRD is a Royal Society University Research Fellow.

\end{document}